\documentclass[%
 aip,
 amsmath,amssymb,
 reprint,%
]{revtex4-1}
\pdfoutput=1
\usepackage{graphicx}
\usepackage{dcolumn}
\usepackage{bm}

\usepackage[utf8]{inputenc}
\usepackage[T1]{fontenc}
\usepackage{mathptmx}
\usepackage{float}
\usepackage{etoolbox}
\usepackage[caption=false]{subfig}

\usepackage[dvipsnames]{xcolor} 

\usepackage{nicefrac}
\usepackage{siunitx} 
\usepackage{nccmath} 
\usepackage{amsmath}
\usepackage[export]{adjustbox}
\usepackage{titlesec}
\usepackage[normalem]{ulem}
\let\Gamma\varGamma

\makeatletter
\def\@email#1#2{%
 \endgroup
 \patchcmd{\titleblock@produce}
  {\frontmatter@RRAPformat}
  {\frontmatter@RRAPformat{\produce@RRAP{*#1\href{mailto:#2}{#2}}}\frontmatter@RRAPformat}
  {}{}
}%
\makeatother

\begin{document}

\preprint{AIP/123-QED}


\title{Photoelectron Transportation Dynamics in GaAs Photocathodes} 



\author{Rui Zhou}
\affiliation{The University of Alabama in Huntsville, Department of Physics \& Astronomy, Huntsville, AL 35899, USA}

\author{Hemang Jani}
\affiliation{The University of Alabama in Huntsville, Department of Physics \& Astronomy, Huntsville, AL 35899, USA}

\author{Yijun Zhang}
\affiliation{School of Electronic and Optical Engineering, Nanjing University of Science and Technology,
Nanjing 210094, China}

\author{Yunsheng Qian}
\affiliation{School of Electronic and Optical Engineering, Nanjing University of Science and Technology,
Nanjing 210094, China}

\author{Lingze Duan}
\email{lingze.duan@uah.edu}
 \affiliation{The University of Alabama in Huntsville, Department of Physics \& Astronomy, Huntsville, AL 35899, USA}

\date{\today}

\begin{abstract}

*The following article has been accepted by \textit{Journal of Applied Physics}.
\newline

\noindent We report here a general theory describing photoelectron transportation dynamics in GaAs semiconductor photocathodes. Gradient doping is incorporated in the model through the inclusion of directional carrier drift. The time-evolution of electron concentration in the active layer upon the injection of an excitation pulse is solved both numerically and analytically. The predictions of the model are compared with experiments via carrier-induced transient reflectivity change, which is measured for gradient-doped and uniform-doped photocathodes using femtosecond pump-probe reflectometry. Excellent agreement is found between the experiments and the theory, leading to the characterization of key device parameters such as diffusion constant and electron decay rates. Comparisons are also made between uniform doping and gradient doping for their characteristics in photoelectron transportation. Doping gradient is found to be able to accelerate electron accumulation on the device surface. These results offer new insights into the dynamics of III-V photocathodes and potentially open a new avenue toward experimental characterization of device parameters.

\end{abstract}


\maketitle 

\section{INTRODUCTION}

Negative-electron-affinity (NEA) III-V semiconductor photocathodes have been widely used in night vision, ultraviolet detection, polarized-electron generation, and photon-enhancement in emission tubes\cite{vv2007semiconductor,la1997photon,liu2008photoemission}. Compared to the traditional metal- or alkali-based photocathodes, III-V photocathodes are able to achieve higher quantum efficiencies (QE)\cite{karkare2014ultrabright,chen2018high}. Recent studies have further indicated that the use of exponential doping structures in the active layer of a III-V photocathode can help enhance the QE\cite{zhang2011photoemission,zou2006gradient,yi2013high}. It has been suggested that such enhancement is due to the built-in electric field caused by the gradient of doping concentration in the active layer. As a result, in such photocathodes, the photoelectrons can be transported toward the surface through both diffusion and directional drift\cite{niu2009influence,yang2007comparison}. In prior studies, theoretical models have been developed to describe the impacts of drift on key specifications of NEA GaAs photocathodes, such as diffusion length and QE\cite{cai2013numerical,niu2009influence}. However, all the existing theories are based on steady-state analysis, which assume the photocathode is under a constant illumination of light.

Meanwhile, an important application of III-V photocathodes is the generation of electron bunches using pulsed lasers\cite{hartmann1999diffusion,honda2013temporal,jin2013picosecond,aulenbacher2002pulse}. Pulsed lasers have also been utilized to investigate the carrier-diffusion dynamics in GaAs photocathodes\cite{hartmann1999diffusion,honda2013temporal,jani2019pre}. The existing steady-state theories are incapable of describing photoelectron transportation in these cases, and a time-dependent, dynamic model has become necessary. Previously, we have developed a diffusion model to explain the behaviors of the photoelectrons in a \textit{uniform-doped} GaAs photocathode following its excitation by a femtosecond laser pulse\cite{jani2019pre,jani2020pump}. In this paper, we generalize the theory to incorporate a built-in electric field and thereby include carrier drift. As such, the model is capable of describing devices with arbitrary doping gradients. We further verify the model by comparing it with experimental results from femtosecond pump-probe reflectometry (PPR) measurements\cite{jani2019pre,jani2020time}. Good experiment-theory agreements are realized, demonstrating the effectiveness of the theory in modeling real devices. Comparisons are also made between uniform doping and gradient doping to examine their differences in electron transportation dynamics, and the impact of doping gradient on surface charge accumulation is discussed. 

\section{\label{sec:2}THEORETICAL MODEL}

Our theory is built upon a two-layer model previously developed for uniform-doped GaAs photocathodes\cite{jani2019pre}. To incorporate directional carrier drift, we first note that, in a gradient-doped semiconductor photocathode, the $p$-type doping concentration varies exponentially with depth (such a device is also known as exponential-doped). Mathematically, this depth-dependent doping profile can be written as 

\begin{equation}
    \label{eq:1}
 N_{d}(x) = N_{d0}\,{\exp(Ax)},
\end{equation}

\noindent where $N_{d0}$ is the doping concentration on the surface of the photocathode, $A$ is the gradient doping coefficient, and $x$ is the depth from the surface. The doping gradient creates a constant electric field normally pointing into the device with a magnitude of

\begin{equation}
   \label{eq:2}
 E = -\frac{d}{dx} \left( \frac{k_{0} T}{q} \ln \frac{N_{d0}}{N_d(x)}\right) = \frac{k_0 TA}{q},  
\end{equation}

\noindent where $k_{0}$ is the Boltzmann constant, $T$ is temperature, and $q$ is the elementary charge. 

\begin{figure}
\centering
\includegraphics[width=0.95\linewidth]{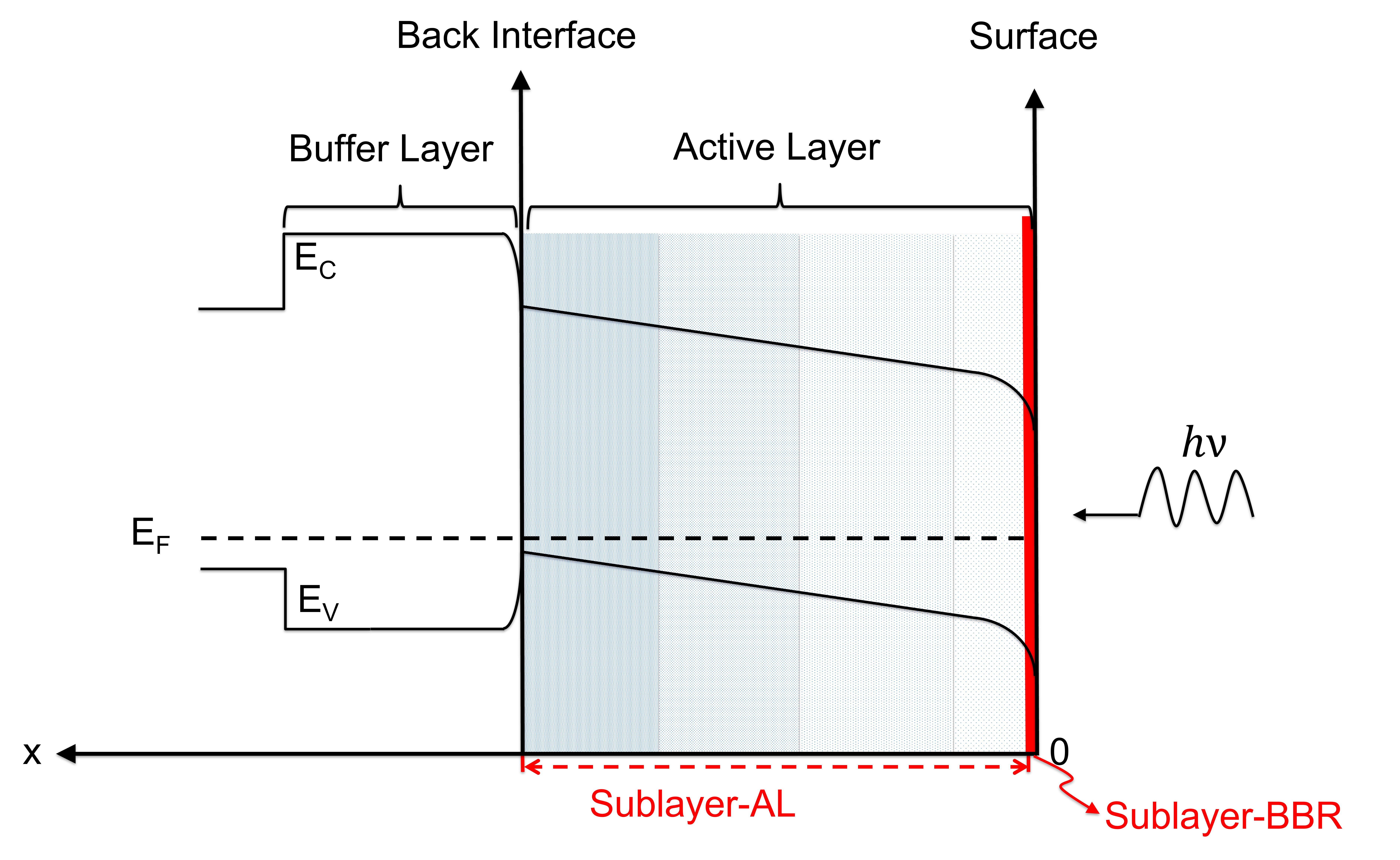}
\caption{The band scheme of a typical gradient-doped GaAs photocathode with illustration of the two sublayers: the active layer (AL) and the band-bending region (BBR).}
\label{fig:1} 
\end{figure}

In the two-layer model, the heavily $p$-doped GaAs layer is divided into two distinct sublayers: an active layer (AL), where most of the photoelectron generation and transportation take place, and a very thin band-bending region (BBR) near the surface, where the photoelectrons accumulate and decay. Separate analyses are applied to AL and BBR respectively based on the different electron behaviors inside them. Fig.~\ref{fig:1} illustrates the band scheme, the coordinate system, the definitions of the sublayers, as well as the doping configuration in the active layer (denser patterns indicate higher doping concentrations).

If we neglect any transverse inhomogeneity in the device and in the optical excitation, the entire system can be described by a one-dimensional (1D) model. Specifically, the 1D electron concentration $n(x,t)$ inside the AL follows the 1D continuity equation, which can be derived from the general continuity equation and written as:

\begin{eqnarray} 
\label{eq:3}
\frac{\partial n(x,t)}{\partial t} &=&D \frac{\partial^2 n(x,t)}{\partial x^2} + \mu_{n} |E| \frac{\partial n(x,t)}{\partial x} + \mu_{n}\,{n(x,t)} \frac{\partial |E|}{\partial x}\nonumber \\
 & &-\frac{n(x,t)}{\tau_{m}} + g_{n},
\end{eqnarray}

\noindent where $D$ is the diffusion coefficient,  $\mu_{n}$ is the electron mobility, $E$ is the doping-induced built-in electric field given by (\ref{eq:2}), $\tau_{m}$ is the mean electron lifetime, and $g_{n}$ is the rate of electron generation caused by all external factors. 

\subsection{The Active Layer (AL)}

As pointed out earlier, the built-in electric field is independent of location in a gradient doping structure. Thus, the term $\partial |E|/{\partial x}=0$ in (\ref{eq:3}). For simplicity, we lump all electron decay into the BBR and treat the AL as decay-free, which eliminates the term ${\Delta n(x,t)}/{\tau_{m}}$  in (\ref{eq:3}). Finally, photoelectron generation by a femtosecond pulse is considered as an instantaneous process, so ${g_{n}}$ can be merged into the initial condition. After these simplifications, the continuity equation becomes

\begin{equation}
    \label{eq:4}
 \frac{\partial n(x,t)}{\partial t} = D \frac{\partial^2 n(x,t)}{\partial x^2}
 +v_{d}\frac{\partial n(x,t)}{\partial x},    
\end{equation}

\noindent where the drift velocity $v_{d}$ is introduced as $v_{d}=\mu_{n}|E|$. The initial condition is $n(x,0)=n_{0}e^{-\alpha x}$, where $\alpha$ is the absorption coefficient in the AL and $n_0$ is a scale factor for the electron population density. For simplicity, $n_{0}$ has been chosen in our model to make $\int_{0}^{d}n_{0}\,e^{-\alpha x}dx=1$, where $d$ is the thickness of the AL. We further assume the BBR acts as an electron “sink” and the back interface of the AL is an impenetrable “wall”. This leads to a Dirichlet boundary condition $n(0,t)=0$ on the AL-BBR interface and a Neumann boundary condition 
${\partial n(x=d,t)}/{\partial x}=0$ on the back interface. 

To solve the differential equation (4), we first notice that $v_{d}$ and $D$ are linked through the Einstein relation ${D}/{\mu_{n}} ={k_{0} T}/{q}$, which results in the relation

\begin{equation}
    \label{eq:5}
 v_{d} = \mu_{n}\left| \frac{k_{0} TA}{q}\right| = DA.   
\end{equation}

\noindent It should be noted here that a "generalized" form of the Einstein relation\cite{kroemer1978einstein} has been considered in our model because the doping concentration in a heavily gradient-doped AL can be comparable or even greater than the effective density of valence-band states in GaAs, which leads to a doping-dependent diffusion coefficient\cite{sze1981physics}. However, further numerical simulations have shown that the impact of such a modification is minimal within the relevant parameter range. Thus, for simplicity, the simple form of the Einstein relation is kept in the model.

The general solution of (\ref{eq:4}) hence is given by

\begin{equation}
    \label{eq:6}
 n(x,t) = 2n_{0}\,{e^{-\gamma x}}\,{\sum_{i=1}^{\infty}} b_{i}\,
 {\sin{(a_{i}x)}}\,{e^{-\beta_{i} Dt}},
\end{equation}

\noindent where $\beta_{i}$ and $\gamma$ have been introduced to simplify the expression and are defined as

\begin{equation}
    \label{eq:7}
 \beta_{i} =\frac{A^2}{4} + {a_{i}}^2,\quad
 \gamma = \frac{A}{2}. 
\end{equation}

\noindent The expansion coefficients $a_{i}$ and $b_{i}$ are ruled by the boundary conditions and the initial condition, respectively. Applying the Neumann boundary condition at the back interface to the general solution (\ref{eq:6}) results in a transcendental equation

\begin{equation}
    \label{eq:8}
 \tan(a_{i} d) = \frac{2}{Ad}(a_{i} d),
\end{equation}

\noindent which gives a set of discrete solutions for $a_{i}$. Eq.~(\ref{eq:8}) has to be solved numerically in general, although analytical solutions are attainable approximately for small $A$ values. This case will be discussed in detail in Section IV. 

Meanwhile, the total number of photoelectrons injected into the BBR can be derived by integrating $n(x,t)$ across the AL and then subtracting the integral from 1,

\begin{equation}
    \label{eq:9}
 N_I(t) = 1- \int_{0}^{d} n(x,t) dx.
\end{equation}

\noindent The electron injection flux from the AL to the BBR is then given by $J(t)={dN_I(t)}/{dt}$.

\begin{figure}
\subfloat[$A=0$ (uniform-doped)]{
\includegraphics[width=0.4827\linewidth]{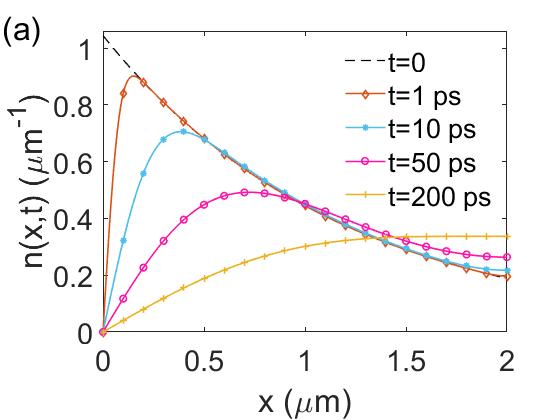} 
}
\subfloat[$A=5$ \si{\micro\metre}\textsuperscript{-1}]{
\includegraphics[width=0.4827\linewidth]{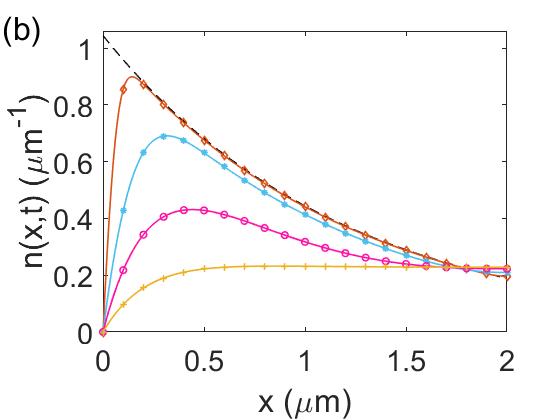}
}
\caption{The time evolution of electron concentration $n(x,t)$ in the sublayer AL for (a) uniform doping ($A=0$) and (b) gradient doping ($A=5$ \si{\micro\metre}\textsuperscript{-1}). Note that $d=2$ \si{\micro\metre} has been assumed.}
\label{fig:2}
\end{figure}

Fig.~\ref{fig:2} shows the time evolution of $n(x,t)$ in an AL of 2 \si{\micro\metre} thickness for (a) $A=0$  (uniform-doped) and (b) $A=5$ \si{\micro\metre}\textsuperscript{-1} (gradient-doped). In both cases, $\alpha=0.85$ \si{\micro\metre}\textsuperscript{-1} has been assumed. Comparing Fig.~\ref{fig:2}(a) and Fig.~\ref{fig:2}(b), it immediately becomes clear that a positive doping gradient pushes the peak concentration further toward the device surface while lowering the $n(x,t)$ profile at a faster pace. Both facts indicate accelerated photoelectron transportation toward the BBR.

\begin{figure}
\subfloat{
\includegraphics[width=0.4827\linewidth]{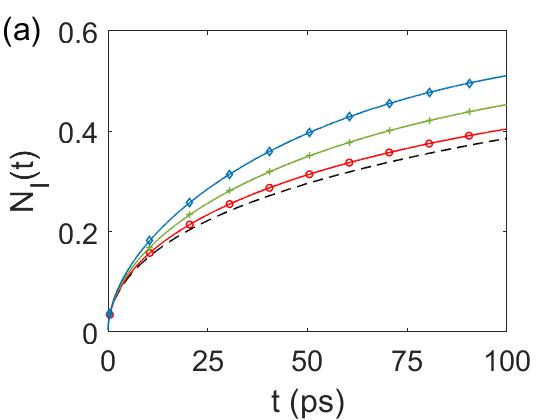} 
}
\subfloat{
\includegraphics[width=0.4827\linewidth]{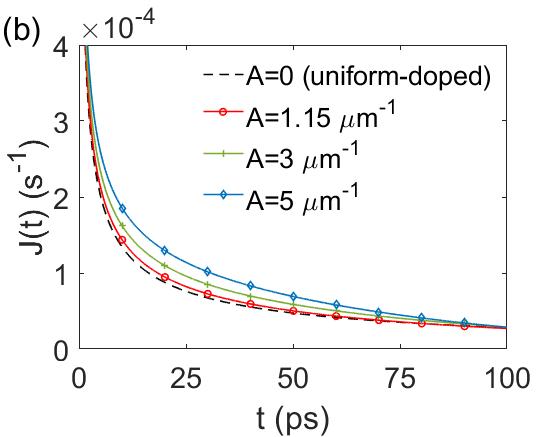}
}
\caption{The impact of doping gradient to photoelectron accumulation on device surface: (a) the growth of the total number of electrons injected from the AL into the BBR for different doping gradients, and (b) the time evolution of the injection flux under various doping gradients ($d=2$ \si{\micro\metre}). Note that $A=0$ indicates uniform doping.}
\label{fig:3}
\end{figure}

The impact of doping gradient can be further evaluated by examining the time evolution of $N_I(t)$ and $J(t)$ with various values of $A$ for a fixed AL thickness $d$. As shown in Fig.~\ref{fig:3}(a), larger doping gradients lead to quicker buildups of the electron population inside the BBR. With $A=5$ \si{\micro\metre}\textsuperscript{-1}, the electron population injected from the AL to the BBR is about 25-30\% higher than it in a uniform-doped device over a time range of 20-100 ps. Meanwhile, the injection flux $J(t)$ is generally 20-50\% higher in a gradient-doped device of $A=5$ \si{\micro\metre}\textsuperscript{-1} than in a uniform-doped device, as shown in Fig.~\ref{fig:3}(b).

\subsection{The Band-Bending Region (BBR)}

In the BBR, all the spatial dependence is neglected due to the infinitesimal thickness. The continuity equation (\ref{eq:3}) reduces to

\begin{equation}
    \label{eq:10}
 \frac{\partial N(t)}{\partial t} = J(t) - \Gamma N(t),
\end{equation}

\noindent where $N(t)$ is the total electron population in the BBR, $J(t)$ is the AL-to-BBR injection flux, and $\Gamma$ is the electron decay rate, which combines all the effects that lead to the reduction of the photoelectron population. $J(t)$ can be found by substituting the general solution of $n(x,t)$ as given by (\ref{eq:6}) into the definition of $N_I(t)$ (\ref{eq:9}). It is then straightforward to solve the differential equation (\ref{eq:10}) to obtain a general solution for $N(t)$,

\begin{eqnarray} 
\label{eq:11}
 N(t) &=& 2{n}_{0}D{\sum_{i=1}^{\infty}}b_{i} 
 \frac{a_{i}-\left[\gamma\sin(a_{i}d)+a_{i}\cos(a_{i}d)\right]e^{-\gamma d}}{\gamma ^2+a_{i}^2}\frac{\beta_i}{\Gamma - \beta_i D} \nonumber\\
 & &\cdot \left(e^{-\beta_i Dt}-e^{-\Gamma t} \right) ,
\end{eqnarray}

\noindent where, again, the coefficients $a_i$ and $b_i$ are determined by the boundary conditions and the initial condition of $n(x,t)$ in the AL.

\begin{figure}
\subfloat{
\includegraphics[width=0.4827\linewidth]{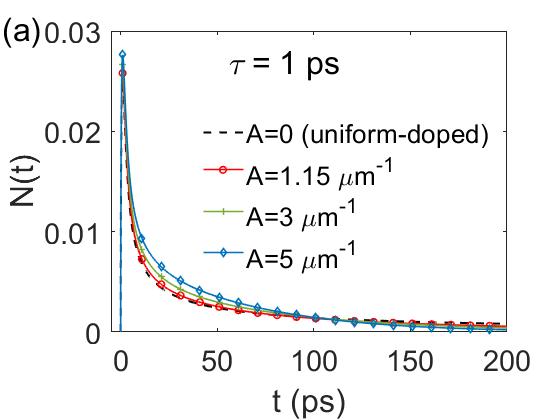} 
}
\subfloat{
\includegraphics[width=0.4827\linewidth]{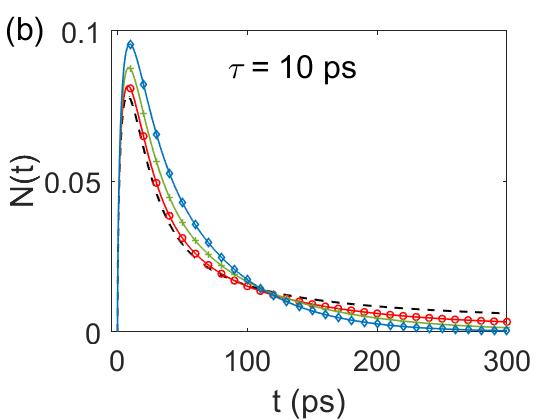}
}\\
\subfloat{
\;\;\includegraphics[width=0.47\linewidth]{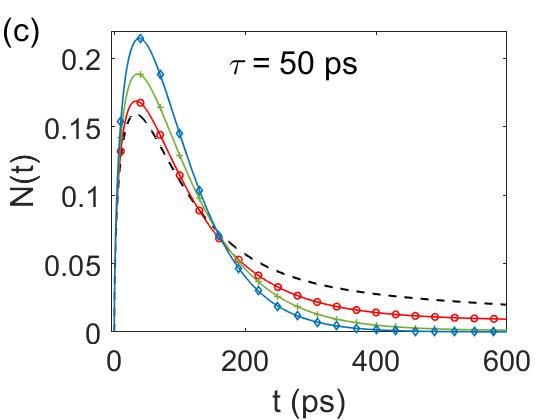}
}
\subfloat{
\includegraphics[width=0.47\linewidth]{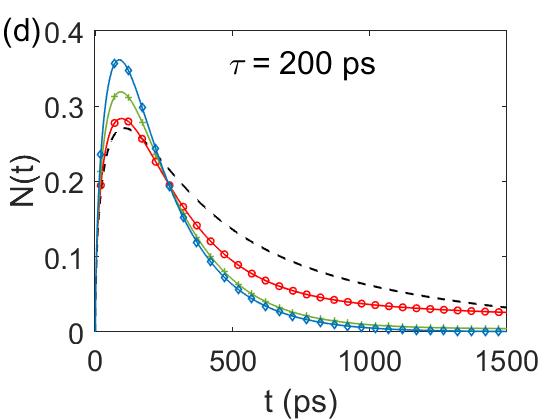}
}
\caption{The injection and decay of the electron population in the BBR leads to a transient behavior of $N(t)$ that features a sharp peak and an exponential tail, as seen here for four different decay lifetimes, $\tau =$ 1, 10, 50, and 200 ps. Uniform doping corresponds to $A=0$.}
\label{fig:4}
\end{figure}

In Fig.~\ref{fig:4}, $N(t)$ is plotted for four different electron decay times, $\tau = 1$ ps, $10$ ps, $50$ ps and $200$ ps, with $\tau$ defined as $\tau=1/\Gamma$. In each case, uniform doping and several gradient-doping cases are plotted to showcase the impact of the doping profile. According to Fig. 4, upon the injection of the laser pulse, the electron population near the device surface first experiences a sharp rise. This is then followed by an exponential population decay. The peak population is influenced by both $A$ and $\tau$. Larger doping gradients generally lead to higher peak populations, especially for large decay times. Such a behavior once again demonstrates the positive impact of the doping gradient on electron accumulation near the device surface.

It should be noted that the 1D total free-electron population $N(t)$ is physically equivalent to the surface charge density in a three-dimensional (3D) picture, which is directly correlated to experimental measurables such as the change of reflectivity. This allows the above theoretical model to be used to explain our femtosecond PPR measurement results as discussed in the next section.

\section{THEORY-EXPERIMENT COMPARISONS}

The above theoretical model can be experimentally verified by means of femtosecond PPR measurement, which probes the transient variation of the surface reflectivity following the injection of an ultrafast laser pulse. According to the well-known Drude theory\cite{tanaka1997subpicosecond,glezer1995laser}, the accumulation of free electrons on the surface of a semiconductor causes a slight change of the surface reflectivity, and the amount of this change is proportional to the area density of the electrons\cite{jani2019pre}, which is directly correlated to the total electron population $N(t)$ in our 1D model as mentioned earlier. Thus, transient evolutions of $N(t)$ such as those shown in Fig.~\ref{fig:4} should be indicative to the behaviors of the transient reflectivity $\Delta R(t)$ measured by the PPR. This allows us to directly compare our theory to experiments. In doing so, not only can we verify the theoretical model, but we are also able to determine key parameters of the tested devices, such as their diffusion coefficients $D$, drift velocities $v_d$, and electron decay times $\tau$. 

Some practical aspects need to be clarified before proper comparisons can be made. In deriving (\ref{eq:11}), we have made the assumption that all the photoelectrons in the BBR share the same decay rate $\Gamma$ (or decay time $\tau$). However, previous studies have shown that multiple decay mechanisms with vastly different decay rates may coexist in actual devices\cite{jani2019pre,hartmann1999diffusion,honda2013temporal}. In particular, a bi-exponential behavior of $\Delta R(t)$ has been found in the GaAs photocathodes studied in this work, indicating two distinctive electron decay rates. To account for the possibility of two decay processes, we generalize our theoretical model by dividing the electron population in the BBR into two independent groups, each following Eq. (\ref{eq:11}) with its own decay rate. The overall transient behavior of $\Delta R(t)$ hence can be modeled by

\begin{equation}
    \label{eq:12}
 \Delta R(t) \propto N(t) = C_1 N_{\Gamma 1}(t) + C_2 N_{\Gamma 2}(t),
\end{equation}

\noindent where $N_{\Gamma 1}(t)$ and $N_{\Gamma 2}(t)$ are the populations of the two electron groups with the decay rates $\Gamma_1$ and $\Gamma_2$, respectively. $C_1$ and $C_2$ represent the partition of the total electron population and satisfy the condition $C_1+C_2=1$. 

Experimental measurement of $\Delta R(t)$ has been performed using a home-built PPR system, which is based on a 6.5-fs Ti:sapphire laser operating at a center wavelength of 800 nm, with an average power of 500 mW and a repetition rate of 83 MHz. More details about the system can be found elsewhere\cite{jani2019pre,jani2020time}. Several samples of GaAs photocathodes have been tested, including two gradient-doped devices, one fabricated with metal organic chemical vapor deposition (MOCVD) and the other fabricated with molecular-beam epitaxy (MBE). The two devices share the same doping structure: a 1-\si{\micro\metre} buffer layer of $p$-Al\textsubscript{0.6}Ga\textsubscript{0.4}As with a doping concentration of $1\times10^{19}$ cm\textsuperscript{-3} directly grown on the $n$-type GaAs substrate, and a 2-\si{\micro\metre} gradient-doped active layer made of $p$-doped GaAs, with a doping concentration changing from $1\times10^{19}$ cm\textsuperscript{-3} near the buffer layer to $1\times10^{18}$ cm\textsuperscript{-3} on the surface. 

\begin{figure}
\subfloat{
\includegraphics[width=0.98\linewidth]{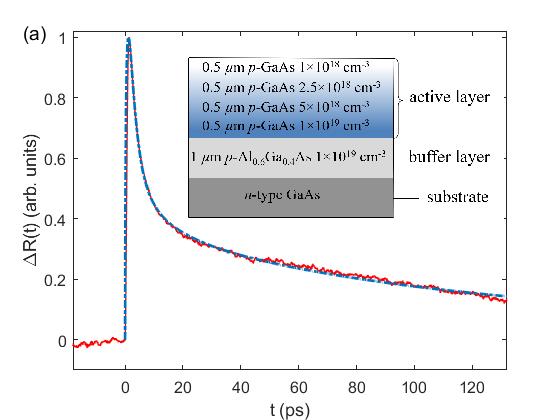} 
}\\
\subfloat{
\includegraphics[width=0.98\linewidth]{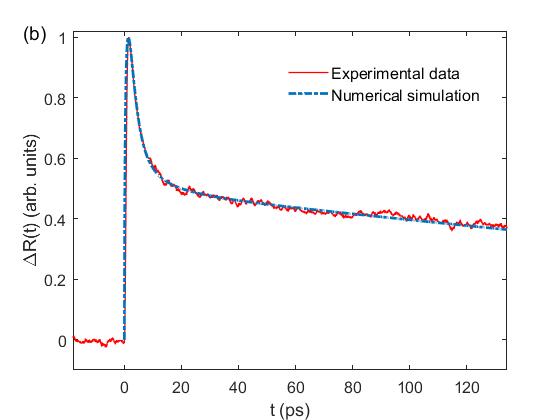}
}
\caption{Comparisons between our theoretical model and experimental data show good agreement for two gradient-doped photocathode samples, fabricated with (a) MOCVD and (b) MBE. Inset: The doping structure of the tested photocathodes.}
\label{fig:5}
\end{figure}

Note that gradient doping is achieved in these devices through the deposition of four uniform-doped sub-layers with progressively decreasing doping concentrations (from bottom up), as shown in the inset of Fig.~\ref{fig:5}(a). Such a stepwise doping structure leads to a ridged electric field profile in the actual devices rather than a constant field as in the theoretical model. Our simulation has shown that the impact of this discrepancy is insignificant in the PPR curve-fitting as the constant-field model is able to capture the average trend of $n(x,t)$ in the AL. This justifies the use of the constant-field model to analyze these devices. 

Fig.~\ref{fig:5}(a) and (b) show the PPR-measured transient reflectivity for the MOCVD and the MBE devices, respectively. In both cases, $\Delta R(t)$ experiences an initial sharp rise followed by a decay process. This general behavior indeed resembles the behavior of $N(t)$ as shown in Fig.~\ref{fig:4}. A closer look at Fig.~\ref{fig:5} further reveals that the decay of $\Delta R(t)$ includes a quick drop immediately following the peak and a long, slowly-decreasing tail, indicating the existence of two decay mechanisms with markedly different decay rates. Using the bi-exponential model (\ref{eq:12}), excellent agreements between theory and experiment are achieved for both devices, as shown in Fig.~\ref{fig:5}. The corresponding fitting parameters are given in Table~\ref{tab:1}. These parameters suggest that the fast decay process has a decay time of about 1 ps, whereas the slow decay process is typically 100 times slower. In both devices, over 90\% of the photoelectrons are lost due to the fast decay process. Although the exact underlying physical mechanisms are not clear solely based on these results, the fast and the slow decay times appear to agree with the typical time scales of surface recombination and photoemission, respectively, according to prior studies on similar GaAs photocathodes
\cite{hang2018temporal,aulenbacher2002pulse,honda2013temporal,jani2019pre}.
\begin{table}
\caption{\label{tab:1}Device parameters used in theoretical model in Fig. 5}
\begin{ruledtabular}
\begin{tabular}{ccccccc}
Sample Type & $D$ (cm\textsuperscript{2}/s) & $v_d$ (cm/s) & $C_1$ & $\tau_1$ (ps) & $C_2$ & $\tau_2$ (ps)\\ 
\hline
MOCVD & 160 & $1.84\times10^6$ & 0.942 & 1.3 & 0.058 & 80\\
MBE & 160 & $1.84\times10^6$ & 0.911 & 1.5 & 0.089 & 180\\
\end{tabular}
\end{ruledtabular}
\end{table} 

It should be pointed out here that the above comparisons between the MOCVD and MBE samples are intended to validate the theoretical model rather than to compare the performances of the photocathodes themselves. Therefore, \textit{normalized} transient reflectivity traces are used and the actual scales of the measured PPR responses are neglected in the current study.

\section{DISCUSSION}

\subsection{Small Doping Gradient}

As mentioned in Section~\ref{sec:2}, the expansion coefficients $a_i$ and $b_i$ in the solution of $n(x,t)$ (\ref{eq:6}) in general cannot be solved analytically due to the transcendental equation (\ref{eq:8}). However, in the special case of a small doping gradient, an approximate analytical solution can be developed. This becomes clear by converting (\ref{eq:8}) into a set of parametric equations with $u=\tan (y)$ and $v=\frac{2}{Ad}y$, where $y=a_i d$, and seeking the intersections between $u(y)$ and $v(y)$. When the slope $\frac{2}{Ad}$ in $v(y)$ is greater than 1, i.e., when $Ad<2$, the intersections are very close to $y=(i-1/2)\pi$, where $i$ is a positive integer. This leads to a set of approximate solutions for $a_i$

\begin{equation}
    \label{eq:13}
 a_i = \frac{2i-1}{2d}\pi, \; i=1,2,3\cdots.
\end{equation}

\noindent Further applying the initial condition $n(x,0)$ to (\ref{eq:6}) and completing the Fourier expansion, the coefficient $b_i$ is found to be

\begin{equation}
    \label{eq:14}
 b_i = \frac{a_i+(-1)^i(\alpha-\gamma)\,e^{-(\alpha-\gamma)d}}{d[(\alpha-\gamma)^2+{a_i}^2]}.
\end{equation}

\noindent Finally, substituting (\ref{eq:13}) and (\ref{eq:14}) into (\ref{eq:11}), a full analytical expression for the total electron population in the BBR (i.e., surface charge density) can be written as

\begin{equation}
    \label{eq:15}
  N(t) = 2n_0 D{\sum_{i=1}^{\infty}}b_i
 \frac{a_i+(-1)^i\gamma \,e^{-\gamma d}}{\gamma ^2+a_i^2}\frac{\beta_i}{\Gamma - \beta_i D}
 \left(e^{-\beta_i Dt}-e^{-\Gamma t} \right).
\end{equation}

\begin{figure}
\subfloat{
\includegraphics[width=0.4827\linewidth]{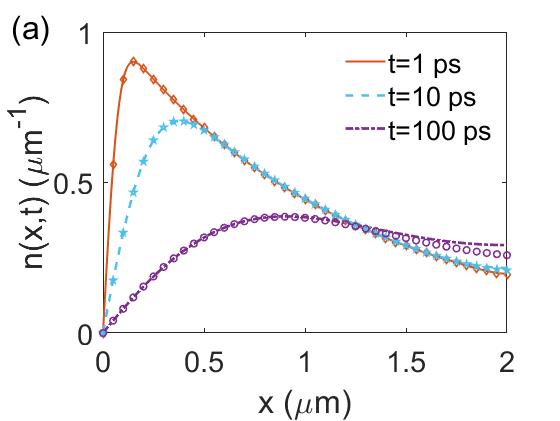} 
}
\subfloat{
\includegraphics[width=0.4827\linewidth]{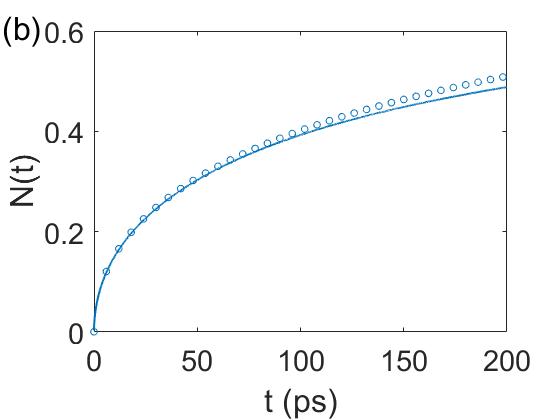}
}
\caption{Comparisons between numerical results (line) and analytical predictions (marker) for (a) $n(x,t)$ and (b) $N(t)$ validate the analytical solution (13)$-$(15) for small doping gradients.}
\label{fig:6}
\end{figure}

Note that, with a typical AL thickness of $d = 2$ \si{\micro\metre}, the solution (\ref{eq:15}) is valid when $A<1$ \si{\micro\metre}\textsuperscript{-1}. To verify the validity of this approximate solution, we have compared the results obtained with the numerical method and the analytical formula for the case of $A=0.5$ \si{\micro\metre}\textsuperscript{-1}. Fig.~\ref{fig:6}(a) 
shows the AL electron concentration $n(x,t)$ versus $x$ at three different delay times. The matching between the numerical results and the analytical predictions is very good for short time scales (e.g., $< 10$ ps), although at longer delay times, slight deviations can be seen near the back interface of the device. Fig.~\ref{fig:6}(b) shows a similar comparison for the BBR electron population $N(t)$. Once again, excellent agreement is achieved between the analytical and the numerical results.

\subsection{Uniform-Doping vs. Gradient-Doping}

Another interesting aspect worthy of further consideration is the comparison between uniform doping and gradient doping. As pointed out in Section~\ref{sec:2} based on the numerical results, gradient doping generally enhances photoelectron transportation from the AL to the BBR, resulting in a faster buildup of the free-electron population on the device surface. But now we can revisit this comparison from a more generic point of view and gain deeper understanding about the underlying physics.

Let us first consider the case of uniform doping. By setting $A=0$ in (\ref{eq:6}) and (\ref{eq:7}), the general solution for $n(x,t)$ in a uniform-doped device can be written as

\begin{equation}
    \label{eq:16}
 n(x,t) = 2n_{0}{\sum_{i=1}^{\infty}} b_{i}\,{\sin{(a_{i}x)}}\,{e^{-{a_i}^2 Dt}},
\end{equation}

\noindent where $a_i$ is given by (\ref{eq:13}) and 

\begin{equation}
    \label{eq:bi}
 b_i=\frac{a_i+(-1)^i\alpha\,e^{-\alpha d}}{d\left(\alpha^2+{a_i}^2\right)},
\end{equation}

\noindent according to (\ref{eq:14}) with $\gamma = 0$. Note that (\ref{eq:16}) is an \textit{exact} solution, and it agrees with the previously reported result based on a diffusion-only model\cite{jani2019pre}.

Now, consider a gradient-doped device. By substituting the parameters in (\ref{eq:7}) into the general solution (\ref{eq:6}) and moving the drift-related terms to the left-hand side of the equation, the following relation is obtained,

\begin{equation}
    \label{eq:17}
 n(x,t)\,e^{\frac{A}{2}\left(x+\frac{1}{2}v_d t\right)} = 2n_{0}{\sum_{i=1}^{\infty}} b_{i}\,{\sin{(a_{i}x)}}\,{e^{-{a_i}^2 Dt}}.
\end{equation}

\noindent If we define the left-hand side as an “effective electron concentration”

\begin{equation}
    \label{eq:18}
 n_{eff}(x,t) = n(x,t)\,e^{\frac{A}{2}\left(x+\frac{1}{2}v_d t\right)},
\end{equation}

\noindent then (\ref{eq:17}) can be rewritten as

\begin{equation}
    \label{eq:19}
 n_{eff}(x,t) = 2n_{0}{\sum_{i=1}^{\infty}} b_{i}\,{\sin{(a_{i}x)}}\,{e^{-{a_i}^2 Dt}}.
\end{equation}

Comparing (\ref{eq:19}) with (\ref{eq:16}), it immediately becomes clear that $n_{eff}(x,t)$ shares the same general solution as the pure diffusive electron concentration in a uniform-doped device, albeit with a different set of initial and boundary conditions, which can be derived according to (\ref{eq:18}) and the initial and boundary conditions for $n(x,t)$. In other words, solving a gradient-doped device in general can be converted into solving a uniform-doping problem for $n_{eff} (x,t)$ and then multiplying the result with $\exp\left[-{\frac{A}{2}\left(x+\frac{1}{2}v_d t\right)}\right]$. This global exponential term can be viewed effectively as a descending exponential envelope propagating at a speed of $\frac{1}{2}v_d$ toward the device surface (i.e., the -$x$ direction). It is the embodiment of the so-called drift-assisted carrier transportation\cite{zhou2020direct}, which accelerates the accumulation of photoelectrons on the device surface. 

\begin{figure}
\centering
\includegraphics[width=0.98\linewidth]{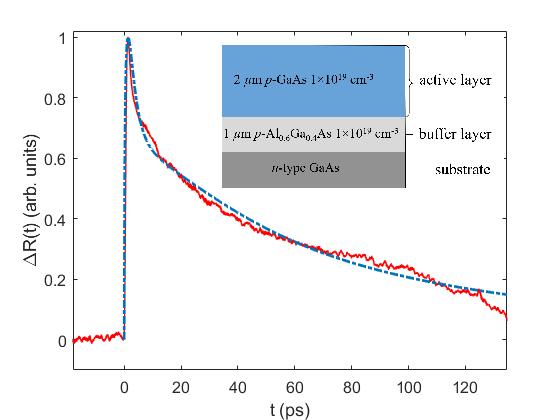}
\caption{A PPR trace (solid) measured with a uniform-doped photocathode shows a good agreement with the theoretical trace (dash-dotted) given by (\ref{eq:15}) and (\ref{eq:12}) with $A=0$. Device parameters used in the theoretical model are: $D=120$ cm\textsuperscript{2}/s, $C_1=0.875$, $\tau_1=1.4$ ps, $C_2=0.125$, $\tau_2=32$ ps. Inset: The doping structure of the tested MOCVD photocathode.}
\label{fig:7} 
\end{figure}

To verify the applicability of our model to uniform-doped devices, we have performed PPR measurements on uniform-doped photocathodes and have compared them with theoretical predictions based on (\ref{eq:15}) and (\ref{eq:12}) under the condition of $A=0$. As shown in Fig.~\ref{fig:7}, good theory-experiment agreement can also be achieved, proving the effectiveness of the model over both uniform and gradient doping profiles.

\section{CONCLUSION}

In conclusion, a general theory describing photoelectron transportation dynamics in GaAs photocathodes has been developed. Time-dependent electron concentration distribution is obtained by solving, both numerically and analytically, a generalized diffusion equation incorporating directional drift caused by gradient doping. Surface charge density is derived to link the theoretical model to experiment via the measurable carrier-induced surface-reflectivity change. The transient reflectivity behaviors of both uniform-doped and gradient-doped GaAs photocathodes are characterized using femtosecond pump-probe reflectometry. Theory-experiment comparisons show excellent agreement, thereby validating the effectiveness of the model in explaining experimental observations. Bi-exponential decay of free-electron population is found in all photocathode samples. The corresponding decay times and partition ratios are derived through curve fitting. Comparisons are made between uniform doping and gradient doping for their characteristics in photoelectron transportation. The impact of doping gradient on the acceleration of electron migration and surface accumulation is discussed based on both numerical results and analytical solutions. Overall, the methodology presented in this paper is complementary to the existing steady-state models and can potentially open a new avenue toward experimental characterization of key device parameters.

\begin{acknowledgments}
This work was funded in part by the National Science Foundation (NSF) under Grants ECCS-1254902 and ECCS-1606836. The authors gratefully acknowledge the support by Dr. Liang Chen and Dr. Shuqin Zhang of the China Jiliang University.
\end{acknowledgments}

\section{DATA AVAILABILITY}
The data that support the findings of this study are available from the corresponding author upon reasonable request.



%
%

%


\nocite{*}
\bibliography{AIPtemplate}

\end{document}